\documentclass[fleqn,usenatbib]{mnras}

\usepackage{newtxtext,newtxmath}
\usepackage[T1]{fontenc}

\DeclareRobustCommand{\VAN}[3]{#2}
\let\VANthebibliography\thebibliography
\def\thebibliography{\DeclareRobustCommand{\VAN}[3]{##3}\VANthebibliography}

\usepackage{graphicx}
\usepackage{amsmath}
\usepackage{booktabs}
\usepackage{bm}
\usepackage{fix-cm}
\usepackage{placeins}

\newcommand{\LCDM}{\ensuremath{\Lambda\mathrm{CDM}}}
\newcommand{\dXe}{\ensuremath{\Delta X_e}}
\newcommand{\ModRec}{{\tt ModRec}}
\newcommand{\fivetwo}{\ensuremath{5\times2}\,pt}
\newcommand{\dchisq}{\ensuremath{\Delta\chi^2}}

\title[Recombination and late-time expansion]{Modified recombination or dynamical dark energy (or both)?}

\author[A. Reeves et al.]{
Alexander Reeves,$^{1}$\thanks{E-mail: alexander.reeves@physics.ox.ac.uk}
Gabriel P. Lynch,$^{2}$
Andrina Nicola,$^{3}$
Jens Chluba,$^{3}$
Alexandre Refregier$^{4}$
\\
$^{1}$Department of Physics, University of Oxford, Denys Wilkinson Building, Keble Road, Oxford OX1 3RH, United Kingdom\\
$^{2}$Department of Physics and Astronomy, University of California, Davis, CA 95616, USA\\
$^{3}$Jodrell Bank Centre for Astrophysics, Department of Physics and Astronomy, The University of Manchester, Manchester M13 9PL, UK\\
$^{4}$Institute for Particle Physics and Astrophysics, ETH Z\"urich, Wolfgang-Pauli-Strasse 27, CH-8093 Z\"urich, Switzerland\\
}

\date{Accepted XXX. Received YYY; in original form ZZZ}
\pubyear{\the\year{}}

\begin{document}
\label{firstpage}
\pagerange{\pageref{firstpage}--\pageref{lastpage}}
\maketitle

\begin{abstract}
Several recent analyses report a hint of modified recombination when the latest cosmic microwave background (CMB) and baryon acoustic oscillation (BAO) data are jointly analysed. We examine whether this preference persists when allowing for departures in the late-time expansion history of the Universe. We fit a seven-parameter reconstruction of the free-electron fraction and a $w_0w_a$ dark-energy model, separately and jointly, to DESI DR2 BAO, a \fivetwo\ combination of galaxy clustering and CMB lensing, and three CMB likelihoods: P-ACT, \textit{Planck} PR4 and \textit{Planck} 2018. With a cosmological constant, our baseline data combination favours early recombination near last scattering, agreeing with previous analyses. Allowing for dynamical dark energy (DDE) removes this preference, while the constraints on $w_0$ and $w_a$ change little when the recombination history is marginalised over. DDE has the lowest Akaike information criterion of the models we consider for every data combination, and adding recombination freedom to it raises the AIC by $5.1$--$11.7$ and is disfavoured by the Bayesian evidence. We interpret these results in terms of CMB predictions for the BAO distances: earlier recombination and DDE both shift these predictions towards the measurements, but only DDE also changes the redshift dependence of the late-time expansion, which anisotropic BAO measurements constrain. We support these conclusions using noiseless mock data, finding that fitting modified recombination to a mock Universe generated with DDE reproduces the apparent preference for modified recombination, whereas DDE cannot similarly reproduce the signatures of modified recombination. The preference for modified recombination found with a cosmological constant therefore reflects the discrepancy between the CMB and BAO, rather than independent evidence for departures from standard recombination physics.
\end{abstract}

\begin{keywords}
cosmological parameters -- cosmic background radiation -- dark energy -- recombination
\end{keywords}

\section{Introduction}

Baryon acoustic oscillation (BAO) measurements from DESI DR2 are in mild tension with the cosmic microwave background (CMB) when both are interpreted within $\Lambda$CDM. This tension is a disagreement between the low-redshift distances that BAO measures, using the sound horizon as a standard ruler, and the values predicted for those distances from the CMB under an assumed model for the late-time expansion. Allowing the dark-energy equation of state to evolve reduces this tension. In addition, dynamical dark energy (DDE) changes the shape of the late-time expansion history, which the BAO measurements probe directly, and these measurements mildly prefer a shape that departs from $\Lambda$CDM. Analysed together, the CMB and BAO favour DDE over $\Lambda$CDM at $3.1\sigma$, and at $2.8$--$4.2\sigma$ when type~Ia supernovae are included, depending on the compilation \citep{DESI:2025zpo}. Abandoning the cosmological constant would be a substantial revision of the standard model, and the claim has accordingly been scrutinised from several directions: whether it is driven by the supernova compilations and their calibration \citep{Efstathiou2024,Colgain2024}, and whether the apparent frequentist significance is matched by Bayesian evidence \citep{Ong:2025utx}. Furthermore, explanations that preserve the cosmological constant and instead reduce the tension in other ways have also been considered.
 
One such possibility is to modify the physics before recombination. Several early-Universe modifications have been proposed in this context. One possibility is to alter the recombination history \citep{LynchB,Mirpoorian:2025rfp}, for example, through varying fundamental constants \citep[e.g.,][]{HartChluba2020,Sekiguchi2021,Schoeneberg2025VFC}, baryon clumping induced by primordial magnetic fields \citep[e.g.,][]{JedamzikPogosian2020,Galli:2021mxk,Thiele2021HT}, or other mechanisms \citep{Rashkovetskyi:2021rwg,Chluba2025ito, Lynch:2026pix}. Earlier recombination can lower the CMB-inferred value of $\omega_{\rm m}r_d^2$, bringing its predictions for BAO distances into closer agreement with observations~\citep{Weiner:2026sfm}. Another possibility explored in the literature is early dark energy (EDE), which introduces additional energy density before recombination, increasing the expansion rate and reducing the sound horizon \citep{Chaussidon:2025npr,Reeves2025}. Both approaches can also accommodate a higher inferred Hubble constant and have therefore been investigated as possible solutions to the tension with local distance-ladder measurements \citep{KnoxMillea2020,DiValentino2021,Schoneberg2022,LynchB,Mirpoorian:2025rfp,Poulin2019,Lee2023,Reeves:2022aoi}.  

Recombination is among the best understood processes in cosmology \citep{Zeldovich68, Peebles1968, Seager2000}, and for the reference scenario, it is now understood to sub-per cent accuracy by dedicated recombination codes \citep{ChlubaThomas2011,HyRec}. It remains interesting to ask how precisely the data themselves can constrain this process, and a phenomenological free-electron history provides that test \citep{Farhang2011, Farhang2013,HartChluba2020,Sekiguchi2021,HartChluba2022,Lee2023}. Using the flexible reconstruction of \citet{LynchA} provided by the \ModRec\ framework, \citet{LynchB} found that earlier recombination improves the fit to \textit{Planck} 2018 and DESI DR1 BAO while retaining a cosmological constant and partially relieving the Hubble tension. ~\citet{Mirpoorian:2025rfp} reached a similar conclusion with DESI DR2 and a different parametrisation.

These analyses leave a pertinent question unresolved: does the preference for modified recombination persist when late-time dark energy is allowed to evolve? Although modified recombination and dynamical dark energy affect different stages of cosmological evolution, their effects on the predicted CMB and BAO statistics can be partially degenerate; for example, both models can modify the distance photons travel during matter domination, which can reduce the discrepancy between CMB and DESI BAO measurements~\citep{Weiner:2026sfm}. The probes nevertheless carry discriminating cosmological information: photon diffusion and the structure of the CMB acoustic peaks constrain recombination \citep{HartChluba2020, LynchA}, while BAO measurements constrain both the absolute distance scale in units of $r_d$ and, through the anisotropic ratio $D_M / D_H$, which is independent of both $r_d$ and $H_0$, the redshift dependence of the late-time expansion. Galaxy clustering and CMB lensing further constrain the matter distribution and its growth, providing an additional test of the cosmologies inferred from the acoustic observables. A joint analysis can therefore establish how the preference for either modification depends on assumptions about the other and which observations distinguish them.

Here we perform this test by fitting a seven-parameter reconstruction of the free-electron history (hereafter dubbed `\ModRec') and an evolving dark-energy model, separately and jointly, to DESI DR2 BAO, three alternative CMB likelihoods, and a \fivetwo\ measurement of galaxy clustering, CMB lensing, and temperature cross-correlations. We analyse the posteriors and model preferences obtained on real data, show how they change as datasets are combined, and calibrate our interpretation with noiseless mock data. Finally, we ask how the conclusions change when the distance-ladder measurement of $H_0$ is taken at face value and included in the analysis.

The remainder of this paper is organised as follows. In Sec.~\ref{sec:setup}, we describe the cosmological models, datasets, and inference methodology. In Sec.~\ref{sec:results}, we present the joint constraints on recombination and dark energy, compare their fits to the data, and use noiseless mocks to investigate the degeneracy between the two scenarios. We discuss our findings and conclude in Sec.~\ref{sec:discussion}.

\section{Models, data and inference}
\label{sec:setup}
 
\subsection{Recombination and dark energy}
\label{sec:models}

We consider four cosmological models: standard \LCDM, modified recombination with a cosmological constant (\ModRec), dynamical dark energy with standard recombination (DDE), and their joint extension (\ModRec$+$DDE). All models share the six base cosmological parameters $\{\omega_b,\omega_{\rm cdm},\sigma_8,n_s,h,\tau\}$, with $\sum m_\nu=0.06\,{\rm eV}$ fixed. The adopted parameter priors are summarised in Table~\ref{tab:priors}. 

\begin{table}
\centering
\caption{Parameter priors adopted in this work. All priors are uniform unless otherwise stated. The $q_i$ are the additive ionization-fraction amplitudes at the seven nodes $z_i$ of the modified-recombination model. For DDE the $w_0w_a$ prior is further restricted to $w_0+w_a<-1/3$.}
\label{tab:priors}
\begin{tabular}{lc}
\hline\hline
Parameter & Prior range \\
\hline
$\omega_b$             & $[0.019,\,0.026]$ \\
$\omega_{\rm cdm}$     & $[0.05,\,0.255]$ \\
$h$                    & $[0.60,\,0.82]$ \\
$\tau$                 & $[0.02,\,0.08]$ \\
$\sigma_8$             & $[0.3,\,1.3]$  \\
$n_s$                  & $[0.84,\,1.1]$ \\
\hline
$w_0$                  & $[-3,\,0.5]$ \\
$w_a$                  & $[-3,\,2]$ \\
\hline
$q_1$ ($z_1=667$)      & $[-0.00117,\,0.00222]$ \\
$q_2$ ($z_2=800$)      & $[-0.00321,\,0.00288]$ \\
$q_3$ ($z_3=933$)      & $[-0.0182,\,0.0248]$ \\
$q_4$ ($z_4=1067$)     & $[-0.0939,\,0.104]$ \\
$q_5$ ($z_5=1200$)     & $[-0.32256,\,0.393]$ \\
$q_6$ ($z_6=1333$)     & $[-0.581,\,0.40]$ \\
$q_7$ ($z_7=1467$)     & $[-0.827,\,0.174]$ \\
\hline
\end{tabular}
\end{table}

\subsubsection{ModRec}
The \ModRec\ model introduces variations in the free electron fraction as a function of redshift that are controlled by seven phenomenological control points. We use the same configuration as in \citet{LynchA}, writing the free electron fraction as
\begin{equation}
X_e(z)\equiv \frac{n_e(z)}{n_{\rm H}(z)}
=X_e^{\rm fid}(z)+\dXe(z),
\end{equation}
where $X_e^{\rm fid}$ is calculated with \textsc{HyRec} \citep{HyRec} in \textsc{class} \citep{CLASS} at each cosmology. The deviation from the standard recombination history, $\Delta X_e(z)$, is defined by seven free amplitudes $q_i$, placed at uniformly spaced redshift nodes between $z=666.7$ and $z=1466.7$ inclusive, with two additional amplitudes fixed to zero at $z=533.3$ and $z=1600$. These fixed boundary amplitudes force the deviation to zero outside of the redshift range $z \in [533.3, 1600]$. In between nodes, $\Delta X_e(z)$ is obtained via cubic spline interpolation of the amplitudes. 

Each amplitude $q_i$ is bounded below by enforcing that the electron fraction at each redshift cannot go below zero and above by $X_e^{\rm max}$, a bound set by the total electron budget available from ionised hydrogen and singly-ionised helium. We neglect the possibility of doubly-ionised helium. In this case, the upper limit is given by $X_e^{\rm max} \equiv 1 + f_{\rm He}$, with $f_{\rm He}=n_{\rm He}/n_{\rm H}\simeq Y_p/[4(1-Y_p)]$. Cubic spline interpolation is not bounds-preserving, so naive interpolation between the nodes is not guaranteed to remain within the physically-allowed range. To account for this, we apply the same non-linear transformation as in \citet{LynchA}. This pushes the amplitudes through a logit mapping such that the bounded domain $[0, X_e^{\rm max}]$ is mapped to an unbounded range, the interpolation is applied between the unbounded images of the free amplitudes, and the inverse transformation is applied across the entire redshift range. This guarantees $X_e(z) \in [0, X_e^{\rm max}]$ at all redshifts. See \citet{LynchA} for details.

During inference, we adopt the uniform priors shown in Table~\ref{tab:priors} for the free amplitudes. The lower bounds for these priors are taken from \citet{LynchA} and represent a 90\% reduction relative to a fixed  $X_e^{\rm fid}$ computed using the \textit{Planck} mean cosmology. The upper limits on these priors are derived from the $5\sigma$ upper limits from the \textit{Planck} 2018 \ModRec\ constraints from \cite{LynchA}. This prior upper bound includes all of the \ModRec\ recombination histories that are consistent with the dataset combinations we consider, which are all at least as constraining as \textit{Planck} 2018.

\subsubsection{DDE}
For DDE, we model departures from a cosmological constant using the Chevallier–Polarski–Linder (CPL) parametrisation of the time-evolution of the dark energy equation of state:
\begin{equation}
    w(a)=w_0+w_a(1-a),
\end{equation}
where $w_0$ is the current dark energy equation of state and $w_a$ describes its linear evolution with scale-factor \citep{Chevallier2001,Linder2003}. We apply  uniform priors $w_0\in[-3,0.5]$ and $w_a\in[-3,2]$. 

\subsection{Data}
\label{sec:data}

\begin{figure*}
\centering
\includegraphics[width=\linewidth]{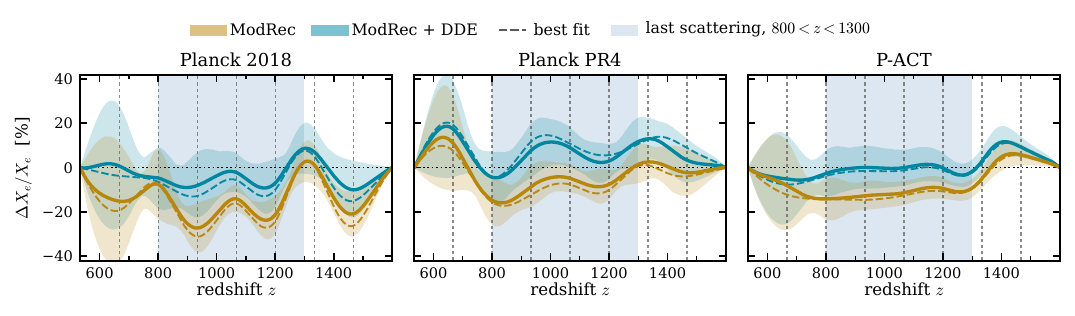}
\\
\caption{Constraints on recombination history for the three baseline CMB combinations, along with BAO and LSS data. Fractional ionization departures for \ModRec\ (gold) and \ModRec$\,+\,$DDE (teal), showing pointwise posterior means and 68 per cent credible bands. Dashed curves indicate best fits, dashed vertical lines mark the nodes, and shading denotes $800<z<1300$.}
\label{fig:joint_reconstruction}
\end{figure*}

\subsubsection{CMB}

We consider three CMB likelihood combinations based on measurements of the temperature and polarization anisotropies.

Our first dataset is the \textit{Planck} 2018 legacy release \citep{Planck2018VI}. We use the foreground-marginalised \texttt{plik\_lite} likelihood for the high-multipole temperature ($TT$), temperature--polarization ($TE$), and polarization ($EE$) power spectra. At low multipoles ($\ell<30$), we include the \textit{Planck} \texttt{Commander} temperature likelihood and the \texttt{SimAll} polarization likelihood, the latter providing a constraint on the optical depth to reionization, $\tau$.

Second, we consider the \textit{Planck} PR4 release, which uses the final reprocessing of the \textit{Planck} observations through the NPIPE pipeline. Compared to the 2018 release, this reprocessing incorporates additional observations collected during satellite repointing manoeuvres, increasing the integration time by approximately $8\%$, alongside improvements in calibration and the treatment of instrumental noise and systematic effects, particularly in polarization. We use the \textsc{HiLLiPoP} high-$\ell$ likelihood, which incorporates a larger sky fraction and an extended multipole range compared to the 2018 analysis, together with the \textsc{LoLLiPoP} low-$\ell$ polarization likelihood \citep{Tristram2024}. For low-$\ell$ temperature, we retain the \textit{Planck} 2018 \texttt{Commander} likelihood.

Finally, we consider measurements from the Atacama Cosmology Telescope (ACT) Data Release 6, which provides high-resolution $TT$, $TE$, and $EE$ power spectra over approximately $10\,000\,\mathrm{deg}^2$ of sky \citep{ACTDR6PS}. The higher angular resolution and polarization sensitivity of ACT provide complementary information to \textit{Planck}, particularly at small angular scales. We use the foreground-marginalised \texttt{ACT-lite} likelihood, supplemented by \textit{Planck} 2018 \texttt{plik\_lite} measurements at $\ell<1000$ in $TT$ and $\ell<600$ in $TE$ and $EE$. These cuts avoid double-counting the correlated small-scale measurements from the two experiments. At low multipoles, we include the \textit{Planck} temperature likelihood and the \textsc{SRoll2} polarization likelihood \citep{Pagano2020}. We refer to this joint combination as P-ACT, following \citet{ACTDR6PS,ACTDR6Params}.

In this work, all likelihoods have been rewritten in \texttt{JAX} to enable rapid, GPU-accelerated analysis\footnote{We release the \texttt{upanda} package containing these \texttt{JAX} reimplementations along with several sampler options and examples showing how to use them \url{https://github.com/alexander-reeves/upanda}}. 

\subsubsection{BAO} We use the baryon acoustic oscillation (BAO) measurements from DESI Data Release 2 (DR2), based on the first three years of observations comprising approximately 14 million galaxies and quasars \citep{DESI2025BAO}. The dataset provides 13 distance measurements at seven effective redshifts spanning $0.295 \leq z_{\rm eff} \leq 2.330$. These are obtained from the Bright Galaxy Survey (BGS), luminous red galaxies (LRGs), emission-line galaxies (ELGs), quasars (QSOs), and the Ly$\alpha$ forest, with the galaxy measurements divided into the BGS, LRG1, LRG2, combined LRG3+ELG1, and ELG2 samples. All samples provide measurements of the transverse comoving distance, $D_M/r_d$, and radial distance, $D_H/r_d = c/(H r_d)$, except BGS, which provides the isotropic distance $D_V/r_d$. We use a \texttt{JAX} reimplementation of the public DESI DR2 likelihood, retaining the full published covariance matrix and consistently computing the sound horizon at the baryon drag epoch, $r_d$, for each cosmological model.

\subsubsection{Multiprobe large-scale structure data} We use the same $5\times2$pt data vector and covariance matrix as \citet{Reeves2025}, combining CMB lensing from ACT DR6 \citep{ACT:2023kun}, galaxy clustering from the DESI Legacy Imaging Surveys \citep{DESI:2022gle}, and the integrated Sachs--Wolfe (ISW) signal from \textit{Planck} 2018 \citep{Planck2018VI}. The DESI Legacy Imaging Survey galaxy sample comprises approximately 12 million luminous red galaxies divided into four tomographic redshift bins. The resulting angular auto- and cross-power spectra provide constraints on the growth of structure and gravitational potentials, complementary to the geometric background information from BAO. We use the full multiprobe covariance matrix of \citet{Reeves2025}, accounting for correlations between the angular power spectra observables. We neglect the cross-covariance with BAO measurements, which was shown to be negligible in~\citet{Reeves2025}.

\subsubsection{Supernovae}

We additionally consider Type Ia supernovae from the DES-Dovekie compilation \citep{Dovekie}, an updated analysis of the DES five-year supernova sample incorporating improved photometric calibration and light-curve modelling. The sample comprises 1623 supernovae from DES and 197 low-redshift supernovae from other surveys. We use a \texttt{JAX} reimplementation of the public likelihood,
marginalising over the supernova absolute magnitude, such that the measurements probe the relative luminosity distances.

\subsection{Inference}
\label{sec:inference}

To make the analysis computationally tractable, we use neural-network emulators trained on \textsc{class} outputs to accelerate the likelihood computation, following a methodology inspired by \citet{CosmoPower} and developed in \citet{Reeves2024,Reeves2025}. The emulators predict CMB spectra up to $\ell=6500$, the \fivetwo\ observables, the BAO observables, and luminosity distances for the supernovae. Importantly for this study, the emulators represent a step beyond what had been previously trained in Ref.~\cite{LynchB}, in that they now include both DDE and \ModRec\ parameters. Their training and validation are described in Appendix~\ref{app:emu}. 

We sample posteriors with \textsc{emcee} \citep{emcee}, using one ensemble of 128 walkers per chain. We assess convergence with the Gelman--Rubin statistic \citep{Gelman:1992zz} computed over 16 groups of pooled walkers, and require $\hat R<1.05$ for every chain. A few walkers in some chains remain stuck in regions more than 50 in $\ln L$ below the maximum, far outside any credible region; we remove their samples, which changes no quoted mean by more than a small fraction of its uncertainty. We obtain best fits by simulated annealing initialized from the maximum a-posteriori points of the respective chains, and Bayesian evidences by nested sampling with \textsc{nautilus} \citep{nautilus}, using five independent seeds to estimate the scatter.

\subsection{CMB--BAO discrepancy}
\label{sec:medi}

In this subsection, we briefly review the CMB--BAO discrepancy observed with DESI DR2 BAO measurements~\citep{DESI2025BAO}, following the interpretation of \citet{Weiner:2026sfm}. In this framework, the discrepancy is expressed as a disagreement between the distance travelled by CMB photons during matter domination inferred directly from CMB and BAO acoustic measurements and that predicted from CMB-calibrated cosmological parameters in the $\Lambda$CDM model. This is a highly informative framing for understanding our results in the next section. The central quantity in this description is the matter-era distance interval (MEDI), $D_M(a_d,a_m)/r_d \equiv D_M(a_d)/r_d - D_M(a_m)/r_d$, defined as the comoving distance interval between some scale factor suitably far into the matter-dominated era, $a_m$, and the baryon drag epoch $a_d$. For convenience, we take $a_m = 1/(1+2.33)$, corresponding to the redshift of the DESI BAO measurement using the Lyman-$\alpha$ forest \citep{DESI:2025zpo}. 

One can estimate the MEDI in two ways. The first is to directly measure it by differencing $D_M(a_m)/r_d$, as measured by DESI DR2 BAO distances, with $D_M(a_d)/r_d$ as determined by the geometric information in the CMB.\footnote{By geometric information, we mean the value of $1/\theta_{\perp}(a_d)$ as inferred by the CMB, marginalised over other parameters. Here, $\theta_{\perp}(a_d)$ is the angular scale of the sound horizon at the end of the baryon drag epoch.} Doing so, \citet{Weiner:2026sfm} finds that the combination of geometric CMB information from Planck, ACT, and SPT-3G and DESI DR2 BAO measurements prefers $D_M(a_d,a_m)/r_d = 55.42 \pm 0.12$. Alternatively, one could predict the expected MEDI, given values of cosmological parameters calibrated with measurements of the CMB power spectra. We may analytically write the scaling of the MEDI as:
\begin{equation}
\frac{D_M(a_d,a_m)}{r_d} \propto
(\omega_{\rm cb}r_d^2)^{-1/2}
\left[1-\Delta_{\rm DE}\right],
\label{eq:medi}
\end{equation}
where we neglect the weak dependence on $a_{\rm eq}$, the scale factor of matter-radiation equality, and the small contribution of massive neutrinos. The factor $\Delta_{\rm DE}$ is the fractional
shortening of the interval by dark energy at $z>2.33$. For an
approximately constant equation of state $w$ at these redshifts,
$\Delta_{\rm DE}\simeq[\rho_{\rm DE}/\rho_m]_{a_m}/(2-12w)$, which is
$\simeq0.5$ per cent for a cosmological constant (i.e. $w=-1$). The $\Lambda$CDM prediction calibrated by the CMB falls below the measured interval by about half a per cent: this difference is the CMB--BAO discrepancy (see Figure~3 of \citet{Weiner:2026sfm}).

Models that modify the early Universe can mediate the CMB--BAO discrepancy to the extent that they raise the CMB-predicted value of the MEDI. From Equation~\ref{eq:medi}, we can identify two routes to doing so: either reducing the value of $\omega_{\rm cb}r_d^2$ inferred by the CMB, thereby raising the MEDI to match the directly measured value, or by removing the suppression from $\Delta_{\rm DE}$, whose magnitude roughly matches the magnitude of the discrepancy. At fixed values of the radiation-to-matter ratio and the baryon-to-photon ratio, each well-determined from the CMB, the former quantity scales as 
\begin{equation}
    \omega_{\rm cb}\,r_d^2 \propto a_\star,
    \label{eq:a_star}
\end{equation}
 where $a_\star$ is the scale factor at last scattering. Earlier recombination lowers $a_\star$ and raises the predicted interval, which is proportional to $a_\star^{-1/2}$. For models that modify the late Universe, on the other hand, suppression due to the dark energy density can also be reduced when the dark energy density at $z>2.33$ is lower than that of a cosmological constant, thereby reducing the discrepancy. This mode of mediation typically occurs in $w_0w_a$ models fit to the CMB+BAO combination and extrapolated to these redshifts.

\section{Results}
\label{sec:results}

\subsection{Joint recombination and dark-energy constraints}
\label{sec:main_results}

In Figure~\ref{fig:joint_reconstruction}, we present the inferred recombination history using our baseline combination of data: CMB + BAO + \fivetwo. At fixed $w=-1$ (i.e., the standard \ModRec\  model with no DDE), all three CMB combinations prefer a lower electron fraction than standard recombination at the four nodes spanning $800\leq z\leq1200$, with the largest shifts for \textit{Planck} 2018. These nodes span the main last-scattering epoch: the standard visibility function peaks at $z\simeq1079$, with 96 per cent of its probability contained in $800<z<1300$. Physically, this suppression in $X_e$ corresponds to earlier recombination, lowering $a_\star$ and, with it, the CMB-inferred $\omega_{\rm cb}r_d^2$ (Equation~\ref{eq:a_star}), which raises the predicted distance interval of equation~(\ref{eq:medi}) towards the value preferred in the data whilst maintaining a good fit to the CMB.

A clear progression is also visible from \textit{Planck} 2018 to \textit{Planck} PR4 and P-ACT: as the CMB measurements become more constraining and probe higher-$\ell$s, the allowed amplitude of the recombination departures decreases, and the reconstructed \(\Delta X_e/X_e\) history becomes smoother (see also Ref.~\cite{ACTDR6Params}). Previous works studying principal component analyses of recombination perturbations \citep{Farhang2011, Farhang2013, Hart2020PCA} revealed that the less-constrained higher order eigenmodes also exhibit more oscillations. Our findings are thus consistent with the high-$\ell$ part of the P-ACT data, allowing us to filter these higher order modes. We confirmed that this statement is independent of the interpolation scheme used to connect the pivot points (see Appendix~\ref{app:interp}).

We now turn to the joint model (\ModRec+DDE) constraints in teal. It can be clearly seen here that allowing \((w_0,w_a)\) to vary simultaneously removes the suppression in $X_e$ observed when fitting \ModRec\ alone, shifting all but one of the node posterior means to within about one standard deviation of zero and mildly broadening their uncertainties. A mild positive departure remains at the highest-redshift ACT node ($\simeq2\sigma$). This node is inversely correlated with the amount of diffusion damping \citep{LynchA}, so the positive shift may accommodate the excess small-scale power measured in the highest multipoles of the P-ACT dataset, relative to the \textit{Planck} 2018 best fit \citep{ACTDR6PS, Poulin:2025nfb}.

\begin{figure}
\centering
\includegraphics[width=\linewidth]{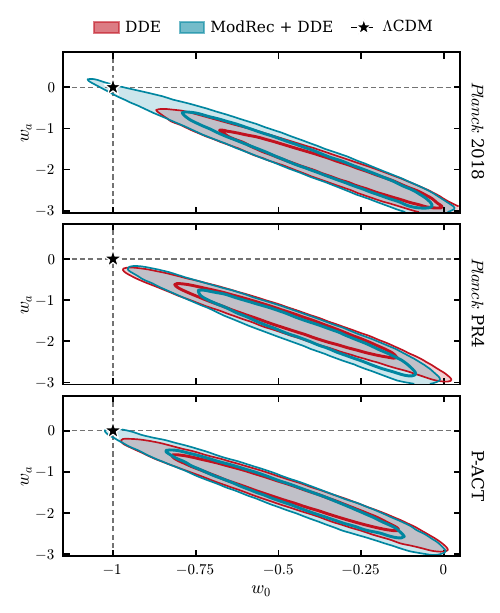}
\\
\caption{Baseline data (CMB+BAO+\fivetwo) constraints in the $w_0w_a$ plane for each CMB likelihood. Crimson contours show DDE and teal contours the joint model; thick and thin lines enclose the 68 and 95 per cent credible regions. The star and dashed lines mark \LCDM. All posteriors extend to the lower edge of the prior, $w_a=-3$.}
\label{fig:w0wa}
\end{figure}

By contrast, the dark-energy constraints remain comparatively stable in the joint model (Figure~\ref{fig:w0wa}): adding the seven recombination amplitudes shifts the posterior means of \(w_0\) and \(w_a\) by at most \(0.42\sigma\) and increases their uncertainties by at most \(26\%\). Thus, the preference for dynamical dark energy in the baseline data combination is largely unchanged after marginalising over substantial freedom in the recombination history, especially once we consider P-ACT. When given the choice to modify the recombination history or the late-time expansion history, the data prefer the latter: both extensions can raise the distance that the CMB predicts between last scattering and $z \approx 2$, but only DDE also changes the anisotropic ratios $D_M / D_H$ at $z<1$, which it does with fewer parameters. We will further analyse the robustness of these statements with targeted mock data in Section~\ref{sec:mocktest}.  

\subsection{Impact of adding supernovae}

\begin{figure}
    \centering
    \includegraphics{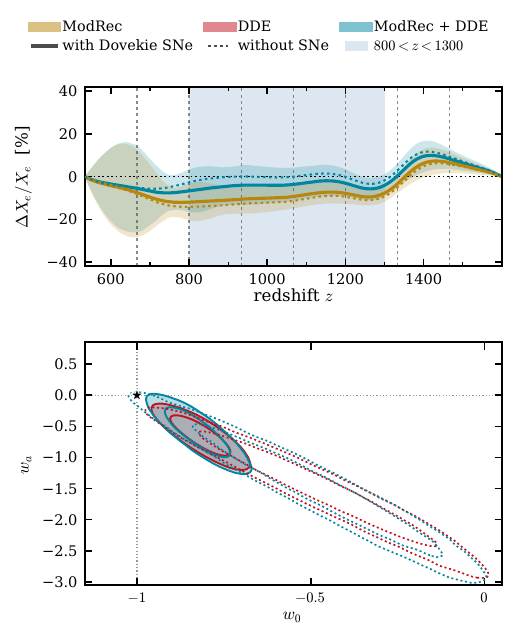}
    \\
    \caption{The impact of adding DES-Dovekie supernovae. Top: fractional ionization departures for \ModRec\ (gold) and \ModRec$\,+\,$DDE (teal), showing pointwise posterior means and 68 per cent credible bands. Dashed curves indicate the posterior means in the baseline data combination, and shading denotes $800<z<1300$. Bottom: 68 and 95 per cent credible regions for DDE (crimson) and the joint model (teal), dashed lines show the baseline data fits and the star marks \LCDM.}
    \label{fig:act_sne}
\end{figure}

Figure~\ref{fig:act_sne} shows the impact of adding the uncalibrated DES-Dovekie supernova sample. Adding DES-Dovekie primarily tightens the dark-energy constraints while leaving the reconstructed recombination history largely unchanged. For P-ACT, the joint-model reconstruction remains consistent with standard recombination within \(1\sigma\) throughout \(800<z<1300\), while the \(w_0\)-\(w_a\) constraints are nearly unchanged by marginalisation over recombination freedom. The other CMB combinations likewise show only minor changes in the reconstructed recombination history after including DES-Dovekie. Therefore, the inclusion of supernovae leaves our main conclusions from section~\ref{sec:main_results} unchanged: the preference for dynamical dark energy is robust to marginalisation over the recombination freedom captured by \ModRec, while the recombination history becomes consistent with the standard prediction once dynamical dark energy is allowed. 

\subsection{Model comparison}
\label{sec:modelcomp}

We compare the models using three statistics: the difference in best-fit $\chi^2$ relative to \LCDM, including contributions from each dataset; the Akaike information criterion, $\Delta\mathrm{AIC}=\Delta\chi^2+2\Delta k$, where $\Delta k$ is the number of additional parameters; and the logarithm of the Bayesian evidence ratio,
\begin{equation}
    \ln B = \ln \frac{Z_{\rm model}}{Z_{\Lambda\mathrm{CDM}}},
    \qquad
    Z = \int d\boldsymbol{\theta}\,
    \mathcal{L}(\boldsymbol{d}|\boldsymbol{\theta})
    \pi(\boldsymbol{\theta}),
\end{equation}
where $\mathcal{L}$ and $\pi$ are the likelihood and parameter prior, respectively. Negative $\Delta\mathrm{AIC}$ and positive $\ln B$ favour the extension over \LCDM.

Among the extensions, DDE is consistently preferred by the AIC across all three CMB datasets, with $\Delta\mathrm{AIC}=-3.6$ to $-10.5$, whereas neither \ModRec\ nor the joint model improves upon \LCDM\ under this criterion (Table~\ref{tab:modelcomp}). The Bayesian evidence, however, is less conclusive for DDE: it is moderately favoured over \LCDM\ for \textit{Planck} 2018 ($\ln B=+2.7$), but slightly disfavoured for PR4 ($-1.0$) and P-ACT ($-0.4$) (see also ~\citet{Ong:2025utx}, which draws similar conclusions with regard to Bayesian evidence and $w_0w_a$). By contrast, models containing \ModRec\ are strongly disfavoured by the Bayesian evidence across all three datasets, with $\ln B\simeq-11$ to $-16$.\footnote{Bayesian evidence depends on the choice of parameter priors, particularly their widths, and the resulting model preferences should therefore be interpreted in this context, especially given the fact that priors are typically not well-motivated for simple $w_0w_a$ models (see e.g. \citep{Efstathiou:2008ed, Hergt:2021qlh}). A systematic investigation of this prior dependence is beyond the scope of this work.}

\begin{figure*}
\centering
\includegraphics[width=\linewidth]{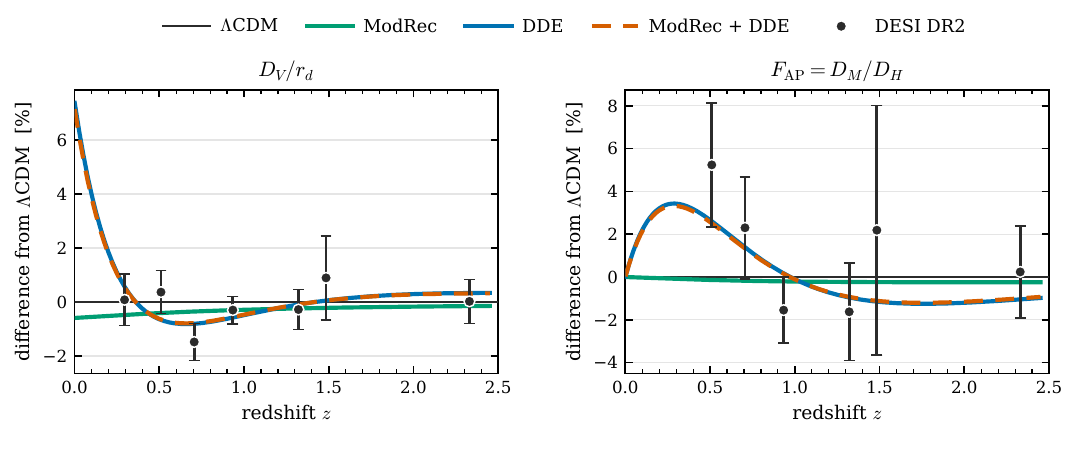}
\\
\caption{The DESI DR2 BAO measurements against the best fit of each model in the baseline combination of data, for P-ACT. Left: the isotropic distance $D_V/r_d$, where $D_V=(zD_M^2D_H)^{1/3}$; right: the anisotropic ratio $F_{\rm AP}=D_M/D_H$, which is independent of both $r_d$ and $H_0$. Both are shown as fractional differences from the \LCDM\ best fit to the same data. At the six redshifts where $D_M$ and $D_H$ are measured separately, $D_V/r_d$ and $F_{\rm AP}$ are formed from the pair, with their marginal errors propagated from the published covariance; the $z=0.295$ bright-galaxy measurement constrains $D_V/r_d$ only and therefore appears only in the left panel.}
\label{fig:bao}
\end{figure*}

The difference between DDE and \ModRec\ can be understood from their contributions to the best-fit $\chi^2$. DDE improves both the CMB ($\Delta\chi^2=-4.1$ to $-7.6$) and BAO ($-3.6$ to $-4.2$) fits relative to \LCDM, whereas \ModRec\ primarily improves the CMB fit ($-6.3$ to $-12.1$), with only a small improvement in BAO ($-0.4$ to $-0.9$). Figure~\ref{fig:bao} illustrates the origin of this difference. Modified recombination lowers $a_\star$, which lowers the CMB-inferred $\omega_{\rm cb}r_d^2$ at a fixed shape of the CMB power spectrum. This allows the model fit to approach the BAO-preferred value of the absolute distance scale of the BAO without distorting the fit to the CMB. However, this model leaves the redshift dependence of the distances within the \LCDM\ family, meaning the fit to the anisotropic BAO is not changed (see $F_{\mathrm{ap}}$ values in Table~\ref{tab:modelcomp}). DDE, on the other hand, changes the late-time expansion history itself, altering the anisotropic ratio $D_M / D_H$ at $z<1$ to be in closer agreement with the DESI DR2 measurements. At the same time, extrapolating the $w_0w_a$ parametrisation to higher redshifts also yields a much smaller dark energy density than in $\Lambda$CDM during the matter-dominated era. This increases the inferred MEDI, allowing DDE to accommodate the BAO-preferred distances while maintaining a good fit to the CMB (see Sec.~\ref{sec:medi}). The joint model follows almost the same trajectory through the BAO data as DDE alone, indicating that additional recombination freedom does not materially change the preferred distance--redshift relation.

Finally, allowing modified recombination in addition to DDE reduces the best-fit $\chi^2$ by $2.3$, $4.9$, and $8.9$ for \textit{Planck} 2018, PR4, and P-ACT, respectively. Under Wilks' theorem (assuming the regularity conditions hold here), if DDE adequately describes the data, the reduction in $\chi^2$ from introducing seven additional parameters is expected to approximately follow a $\chi^2$ distribution with seven degrees of freedom. Even the largest improvement in the fit, obtained for P-ACT, corresponds to a $p$-value of approximately $0.26$, and therefore does not provide significant evidence for additional recombination freedom. Thus, neither the improvement in best-fit $\chi^2$ nor the model comparison statistics provide evidence that additional recombination freedom is warranted by the data.

\begin{table*}
\centering
\caption{Baseline model comparison relative to \LCDM. For each CMB dataset in the baseline combination of data the first column shows the number of additional parameters beyond $\Lambda$CDM, the next three columns give the $\Delta \chi^2$ for each component of the likelihood, the following two split the BAO contribution into its $D_V/r_d$ and $F_{\rm AP}=D_M/D_H$ parts, followed by the total $\Delta \chi^2$. The $D_V/r_d$ and $F_{\rm AP}$ columns give the $\Delta\chi^2$ of each set of measurements on its own, using their marginal uncertainties (the error bars in Fig.~\ref{fig:bao}). Because $D_V/r_d$ and $F_{\rm AP}$ measured at the same redshift are correlated, most strongly for the Ly$\alpha$ forest at $z=2.33$, these two columns need not sum to the BAO column, which gives the contribution of the full DESI DR2 likelihood. The final two columns tabulate the $\Delta \mathrm{AIC}=\Delta \chi^2+2\Delta N_{\rm par}$ and the Bayesian evidence respectively (the latter is computed from nested sampling outputs).}
\label{tab:modelcomp}
\setlength{\tabcolsep}{5pt}
\begin{tabular}{llrrrrrrrrr}
\toprule
 & & & \multicolumn{5}{c}{$\dchisq$ by component} & & & \\
\cmidrule(lr){4-8}
CMB data & Model & $\Delta k$ & CMB & \fivetwo & BAO & $D_V/r_d$ & $F_{\rm AP}$ & $\dchisq$ & $\Delta$AIC & $\ln B$ \\
\midrule
\textit{Planck} 2018 & \ModRec & 7 & $-9.8$ & $-2.3$ & $-0.7$ & $-0.8$ & $0.0$ & $-12.8$ & $+1.2$ & $-11.3$ \\
 & DDE & 2 & $-7.6$ & $-2.6$ & $-4.2$ & $-1.6$ & $-3.7$ & $-14.5$ & $-10.5$ & $+2.7$ \\
 & \ModRec$+$DDE & 9 & $-9.6$ & $-2.7$ & $-4.4$ & $-1.8$ & $-3.2$ & $-16.8$ & $+1.2$ & $-13.2$ \\
\midrule
\textit{Planck} PR4 & \ModRec & 7 & $-6.3$ & $-0.3$ & $-0.4$ & $-0.4$ & $0.0$ & $-7.1$ & $+6.9$ & $-14.8$ \\
 & DDE & 2 & $-4.1$ & $0.0$ & $-3.6$ & $-1.2$ & $-3.0$ & $-7.6$ & $-3.6$ & $-1.0$ \\
 & \ModRec$+$DDE & 9 & $-8.6$ & $-0.3$ & $-3.6$ & $-1.0$ & $-3.5$ & $-12.5$ & $+5.5$ & $-16.1$ \\
\midrule
P-ACT & \ModRec & 7 & $-12.1$ & $-0.4$ & $-0.9$ & $-0.9$ & $0.0$ & $-13.3$ & $+0.7$ & $-14.1$ \\
 & DDE & 2 & $-5.1$ & $0.0$ & $-4.0$ & $-1.7$ & $-3.1$ & $-9.1$ & $-5.1$ & $-0.4$ \\
 & \ModRec$+$DDE & 9 & $-13.8$ & $-0.1$ & $-4.1$ & $-1.8$ & $-3.1$ & $-18.0$ & $0.0$ & $-16.2$ \\
\bottomrule
\end{tabular}

\end{table*}

\subsection{Controlled tests with mock data}
\label{sec:mocktest}

\begin{figure*}
\centering
\includegraphics[width=\linewidth]{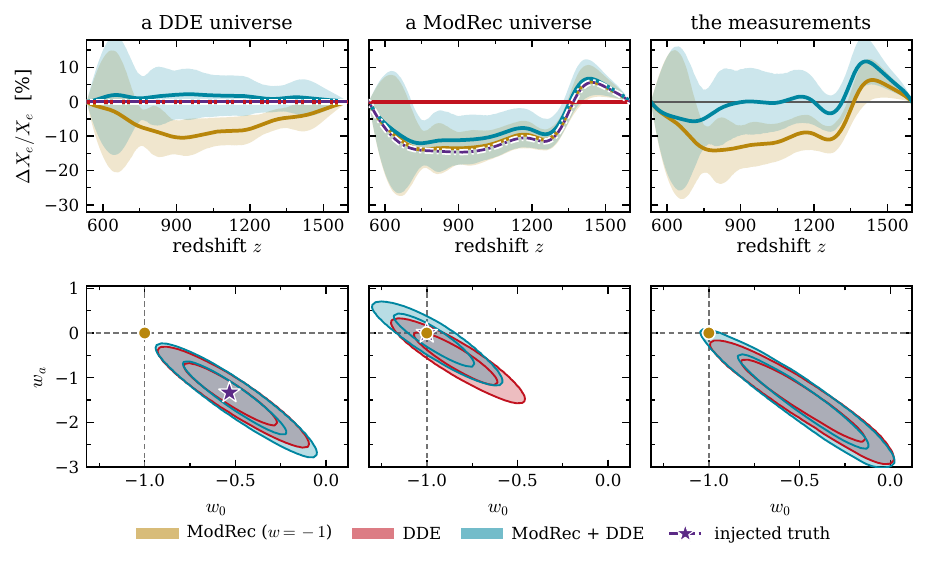}
\caption{Controlled tests with noiseless mock data. The columns show a mock generated from the best-fit DDE model with standard recombination (left), a mock generated from the best-fit \ModRec\ model with $w=-1$ (middle), and the real measurements for comparison (right). Colours indicate the model being fitted: \ModRec\ (gold), DDE (crimson), and their joint extension (teal). Top: reconstructed ionization histories, with the standard recombination history (solid grey) and the injected mock histories (dash-dotted violet). Bottom: 68 and 95 per cent credible regions in the $(w_0,w_a)$ plane. Violet stars mark the injected values and the gold dot marks \LCDM. All mocks use the observational covariances and fiducial nuisance parameters described in the text.}
\label{fig:mocktest}
\end{figure*}

The preceding results show that allowing for DDE removes the apparent preference for modified recombination, while the DDE constraints remain largely unchanged when recombination freedom is introduced (Fig.~\ref{fig:joint_reconstruction}). To understand this behaviour, we perform two controlled tests: whether a genuine DDE signal can be mistaken for modified recombination, and whether a genuine recombination departure can instead be absorbed by DDE. We generate noiseless mock CMB, BAO, and \fivetwo\ data from the real-data best fits of each model, retaining the observational covariances and fixing nuisance parameters to the centres of their Gaussian priors. Figure~\ref{fig:mocktest} shows the results for P-ACT, while Table~\ref{tab:mock_chi2} gives the best-fit mismatches for all three CMB combinations.

We first consider a mock universe with DDE and standard recombination (left column of Fig.~\ref{fig:mocktest}). When fitted with \ModRec\ alone, the mock produces an apparent suppression of the ionization fraction resembling that found in the real data, despite having no injected recombination modification. However, \ModRec\ cannot fully reproduce the mock: for P-ACT, the best-fit mismatch is $\Delta\chi^2=4.58$, of which $3.82$ comes from BAO. The same pattern holds for the other CMB combinations (Table~\ref{tab:mock_chi2}). Thus, modified recombination can reproduce the distance scale of the DDE mock, but not the anisotropic ratios $D_M / D_H$ that DDE imprints at $z<1$.

We next consider the reverse test, generating a mock with modified recombination and $w=-1$ (middle column of Fig.~\ref{fig:mocktest}). The joint model recovers the injected ionization history, while the dark-energy constraints remain consistent with a cosmological constant. Fitting DDE alone also yields constraints consistent with $w=-1$ at $1\sigma$, rather than generating a spurious preference for evolving dark energy. Moreover, DDE provides a substantially worse fit to the mock, with $\Delta\chi^2=9.94$ for P-ACT, dominated by the CMB contribution of $8.88$. This behaviour is again seen across all three CMB combinations. Unlike the previous test, the mismatch arises primarily because changes to late-time expansion cannot reproduce the CMB signatures of modified recombination.

These tests demonstrate that the two extensions are partially degenerate but not interchangeable. In particular, an apparent recombination departure can arise when DDE is incorrectly fixed to a cosmological constant, whereas the injected recombination departures considered here are not erased by allowing DDE. This provides an explanation for the real-data results: the preference for DDE persists after marginalizing over recombination freedom, while the reconstructed ionization history becomes consistent with standard recombination. Together with the model comparison results in Sec.~\ref{sec:modelcomp}, the mocks support the interpretation that the apparent recombination signal in the restricted \ModRec\ model with a cosmological constant is largely compensating for an altered expansion history after recombination preferred by the data. 

\begin{table*}
\centering
\caption{Best-fit $\Delta\chi^2$ when fitting each noiseless mock with the competing model, relative to the generating model. The CMB contribution combines the high- and low-multipole likelihoods. Totals include the small prior contributions (not shown), and may differ slightly from the sum of the displayed components due to rounding.}
\label{tab:mock_chi2}
\begin{tabular}{llrrrr}
\hline
CMB dataset & Mock $\rightarrow$ fit
& CMB & BAO & \fivetwo & Total \\
\hline
ACT DR6      & DDE $\rightarrow$ \ModRec
             & 0.50 & 3.82 & 0.20 & 4.58 \\
Planck PR4   & DDE $\rightarrow$ \ModRec
             & 0.21 & 5.76 & 0.17 & 6.17 \\
Planck 2018  & DDE $\rightarrow$ \ModRec
             & 0.56 & 7.13 & 0.39 & 8.11 \\
\hline
ACT DR6      & \ModRec $\rightarrow$ DDE
             & 8.88 & 0.87 & 0.16 & 9.94 \\
Planck PR4   & \ModRec $\rightarrow$ DDE
             & 4.95 & 0.64 & 0.09 & 5.69 \\
Planck 2018  & \ModRec $\rightarrow$ DDE
             & 5.77 & 2.76 & 0.22 & 8.77 \\
\hline
\end{tabular}
\end{table*}

\section{Discussion and conclusions}
\label{sec:discussion}
Previous studies demonstrated that modified recombination can improve the fit to current CMB and BAO data while retaining a cosmological constant. However, these analyses assume a fixed late-time equation of state, leaving open the possibility that the results could change once the late-time expansion history is allowed to vary. We investigate this possibility by allowing both modified recombination and dynamical dark energy (DDE), separately and jointly, in fits to CMB and DESI DR2 BAO data. When both extensions are allowed, the data favour DDE: the suppression of the ionization fraction inferred at fixed $w=-1$ largely disappears, with the recombination history becoming consistent with the standard prediction. Meanwhile, the constraints on $(w_0,w_a)$ remain largely unchanged when recombination freedom is introduced. This conclusion holds across the CMB datasets considered and is unchanged after adding \fivetwo\ constraints and uncalibrated Type-Ia supernovae data.

The origin of this behaviour can be traced to the DESI BAO distances. Modified recombination lowers $a_\star$, the scale factor at last scattering, which brings the $\omega_{\rm cb}r_d^2$ inferred by the CMB into agreement with the DESI BAO preference while preserving the \LCDM\ redshift dependence of the distances. DDE, instead, changes that redshift dependence directly. As a result, \ModRec\ cannot reproduce the redshift dependence preferred by DESI, even though it can mimic part of the CMB response of DDE. Our noiseless mock tests confirm this interpretation. Data generated with DDE and standard recombination produce an apparent ionization suppression when fitted with \ModRec\ alone, but the remaining mismatch is dominated by BAO. Conversely, a mock generated with modified recombination does not produce a spurious preference for DDE, and the mismatch in this case is dominated by the CMB. These tests identify the anisotropic DESI ratios $D_M / D_H$ at $z<1$ as the key discriminator.

The model comparison further supports this interpretation. Once DDE is included, allowing seven additional recombination parameters produces only modest improvements in the best-fit $\chi^2$, and both AIC and Bayesian evidence favour DDE over the joint extension. The baseline data therefore provide no statistical support for additional recombination freedom beyond DDE. The preference for DDE over \LCDM\ itself is less conclusive, however, and depends on the CMB likelihood and model-comparison statistic.

Even if additional photon transport effects \citep{Chluba2025ito, Lynch:2026pix} and more general consideration of reionization \citep{Sailer2026, Weiner:2026sfm} may slightly modify the conclusions obtained within the \ModRec\ treatment presented here, our results do put pressure on modified recombination as an explanation for the Hubble tension when DDE is allowed.

Upcoming CMB observations from the Simons Observatory \citep{Abitbol:2025SO} will provide more precise tests of recombination physics, while further measurements from DESI \citep{DESI:2016fyo}, Euclid \citep{Mellier:2024Euclid}, and Rubin--LSST \citep{Ivezic:2008LSST} will improve constraints on the late-time expansion and growth histories. Together, these observations will help establish whether the DESI preference for DDE persists and further distinguish early- and late-time departures from \LCDM.

\section*{Acknowledgements}
Computations were performed on the Euler cluster at ETH Z\"urich. AR acknowledges support from the Swiss National Science Foundation (SNSF) through a Postdoc.Mobility fellowship (grant no. P500-2\_242222). GPL was supported by DOE Office of Science award DESC0009999 and the Michael and Ester Vaida Endowed Chair in Cosmology and Astrophysics.  AN acknowledges support from the European Research Council (ERC) under the European Union’s Horizon 2020 research and innovation program with Grant agreement No. 101163128. 
AI use statement: Anthropic’s Claude was used to assist with the implementation of code for producing figures and for writing suggestions. All AI-assisted code and text were reviewed by the authors, who take full responsibility for the entire content of this work.

\section*{Data Availability}
Chains, emulators, and emulator training data will be made public upon publication. All likelihoods and datasets used are public.

\bibliographystyle{mnras}
\bibliography{bib}

@ARTICLE{Sailer2026,
       author = {{Sailer}, Noah and {Farren}, Gerrit S. and {Ferraro}, Simone and {White}, Martin},
        title = "{Addressing Tensions in {\ensuremath{\Lambda}}CDM Cosmology by an Increase in the Optical Depth to Reionization}",
      journal = {\prl},
         year = 2026,
        month = feb,
       volume = {136},
          eid = {081002},
        pages = {081002},
          doi = {10.1103/6r54-8lv4},
archivePrefix = {arXiv},
       eprint = {2504.16932},
 primaryClass = {astro-ph.CO},
       adsurl = {https://ui.adsabs.harvard.edu/abs/2026PhRvL.136h1002S}
}

@ARTICLE{Chluba2025ito,
       author = {{Chluba}, Jens and {Vasil}, Geoffrey and {Battye}, Richard},
        title = "{It{\^o} calculus meets the Hubble tension: effects of small-scale electron density fluctuations on the CMB anisotropies}",
      journal = {\mnras},
         year = 2025,
        month = oct,
       volume = {543},
       number = {1},
        pages = {456-478},
          doi = {10.1093/mnras/staf1465},
archivePrefix = {arXiv},
       eprint = {2505.22242},
 primaryClass = {astro-ph.CO},
       adsurl = {https://ui.adsabs.harvard.edu/abs/2025MNRAS.543..456C}
}

@ARTICLE{Hart2020PCA,
       author = {{Hart}, Luke and {Chluba}, Jens},
        title = "{Improved model-independent constraints on the recombination era and development of a direct projection method}",
      journal = {\mnras},
         year = 2020,
        month = jul,
       volume = {495},
       number = {4},
        pages = {4210-4226},
          doi = {10.1093/mnras/staa1426},
archivePrefix = {arXiv},
       eprint = {1912.04682},
 primaryClass = {astro-ph.CO},
       adsurl = {https://ui.adsabs.harvard.edu/abs/2020MNRAS.495.4210H}
}

@article{Efstathiou:2008ed,
    author = "Efstathiou, G.",
    title = "{Limitations of Bayesian Evidence Applied to Cosmology}",
    eprint = "0802.3185",
    archivePrefix = "arXiv",
    primaryClass = "astro-ph",
    doi = "10.1111/j.1365-2966.2008.13498.x",
    journal = "Mon. Not. Roy. Astron. Soc.",
    volume = "388",
    pages = "1314",
    year = "2008"
}

@article{Hergt:2021qlh,
    author = "Hergt, Lukas T. and Handley, Will J. and Hobson, Michael P. and Lasenby, Anthony N.",
    title = "{Bayesian evidence for the tensor-to-scalar ratio $r$ and neutrino masses $m_\nu$: Effects of uniform vs logarithmic priors}",
    eprint = "2102.11511",
    archivePrefix = "arXiv",
    primaryClass = "astro-ph.CO",
    doi = "10.1103/PhysRevD.103.123511",
    journal = "Phys. Rev. D",
    volume = "103",
    pages = "123511",
    year = "2021"
}

@article{ACT:2023kun,
    author = "Madhavacheril, Mathew S. and others",
    collaboration = "ACT",
    title = "{The Atacama Cosmology Telescope: DR6 Gravitational Lensing Map and Cosmological Parameters}",
    eprint = "2304.05203",
    archivePrefix = "arXiv",
    primaryClass = "astro-ph.CO",
    reportNumber = "FERMILAB-PUB-23-206-PPD",
    doi = "10.3847/1538-4357/acff5f",
    journal = "Astrophys. J.",
    volume = "962",
    number = "2",
    pages = "113",
    year = "2024"
}

@ARTICLE{Thiele2021HT,
       author = {{Thiele}, Leander and {Guan}, Yilun and {Hill}, J. Colin and {Kosowsky}, Arthur and {Spergel}, David N.},
        title = "{Can small-scale baryon inhomogeneities resolve the Hubble tension? An investigation with ACT DR4}",
      journal = {\prd},
         year = 2021,
        month = sep,
       volume = {104},
       number = {6},
          eid = {063535},
        pages = {063535},
          doi = {10.1103/PhysRevD.104.063535},
archivePrefix = {arXiv},
       eprint = {2105.03003},
 primaryClass = {astro-ph.CO},
       adsurl = {https://ui.adsabs.harvard.edu/abs/2021PhRvD.104f3535T}
}

@article{Galli:2021mxk,
    author = "Galli, Silvia and Pogosian, Levon and Jedamzik, Karsten and Balkenhol, Lennart",
    title = "{Consistency of Planck, ACT, and SPT constraints on magnetically assisted recombination and forecasts for future experiments}",
    eprint = "2109.03816",
    archivePrefix = "arXiv",
    primaryClass = "astro-ph.CO",
    doi = "10.1103/PhysRevD.105.023513",
    journal = "Phys. Rev. D",
    volume = "105",
    number = "2",
    pages = "023513",
    year = "2022"
}

@article{DESI:2022gle,
    author = "Zhou, Rongpu and others",
    collaboration = "DESI",
    title = "{Target Selection and Validation of DESI Luminous Red Galaxies}",
    eprint = "2208.08515",
    archivePrefix = "arXiv",
    primaryClass = "astro-ph.CO",
    doi = "10.3847/1538-3881/aca5fb",
    journal = "Astron. J.",
    volume = "165",
    number = "2",
    pages = "58",
    year = "2023"
}

@article{Abitbol:2025SO,
    author        = "Abitbol, M. and others",
    collaboration = "Simons Observatory",
    title         = "{The Simons Observatory: Science Goals and Forecasts for the Enhanced Large Aperture Telescope}",
    eprint        = "2503.00636",
    archivePrefix = "arXiv",
    primaryClass  = "astro-ph.IM",
    doi           = "10.1088/1475-7516/2025/08/034",
    journal       = "JCAP",
    volume        = "08",
    pages         = "034",
    year          = "2025"
}

@article{DESI:2016fyo,
    author = "Aghamousa, Amir and others",
    collaboration = "DESI",
    title = "{The DESI Experiment Part I: Science, Targeting, and Survey Design}",
    journal = "ArXiv e-prints",
    eprint = "1611.00036",
    archivePrefix = "arXiv",
    primaryClass = "astro-ph.IM",
    reportNumber = "FERMILAB-PUB-16-517-AE",
    year = "2016"
}

@article{Mellier:2024Euclid,
    author        = "Mellier, Y. and others",
    collaboration = "Euclid",
    title         = "{Euclid. I. Overview of the Euclid Mission}",
    eprint        = "2405.13491",
    archivePrefix = "arXiv",
    primaryClass  = "astro-ph.CO",
    doi           = "10.1051/0004-6361/202450810",
    journal       = "Astron. Astrophys.",
    volume        = "697",
    pages         = "A1",
    year          = "2025"
}

@article{Ivezic:2008LSST,
    author        = "Ivezi{\'c}, {\v Z}eljko and others",
    collaboration = "LSST",
    title         = "{LSST: From Science Drivers to Reference Design and Anticipated Data Products}",
    eprint        = "0805.2366",
    archivePrefix = "arXiv",
    primaryClass  = "astro-ph",
    doi           = "10.3847/1538-4357/ab042c",
    journal       = "Astrophys. J.",
    volume        = "873",
    number        = "2",
    pages         = "111",
    year          = "2019"
}

@ARTICLE{Schoeneberg2025VFC,
       author = {{Sch{\"o}neberg}, Nils and {Vacher}, L{\'e}o},
        title = "{The mass effect {\textemdash} variations of the electron mass and their impact on cosmology}",
      journal = {\jcap},
         year = 2025,
        month = mar,
       volume = {2025},
       number = {3},
          eid = {004},
        pages = {004},
          doi = {10.1088/1475-7516/2025/03/004},
archivePrefix = {arXiv},
       eprint = {2407.16845},
 primaryClass = {astro-ph.CO},
       adsurl = {https://ui.adsabs.harvard.edu/abs/2025JCAP...03..004S}
}

@ARTICLE{Farhang2011,
   author = {{Farhang}, M. and {Bond}, J.~R. and {Chluba}, J.},
    title = "{Semi-blind Eigen Analyses of Recombination Histories Using Cosmic Microwave Background Data}",
  journal = {\apj},
archivePrefix = "arXiv",
   eprint = {1110.4608},
 primaryClass = "astro-ph.CO",
     year = 2012,
    month = jun,
   volume = 752,
      eid = {88},
    pages = {88},
      doi = {10.1088/0004-637X/752/2/88},
   adsurl = {http://adsabs.harvard.edu/abs/2012ApJ...752...88F}
}

@ARTICLE{Farhang2013,
   author = {{Farhang}, M. and {Bond}, J.~R. and {Chluba}, J. and {Switzer}, E.~R.
	},
    title = "{Constraints on Perturbations to the Recombination History from Measurements of the Cosmic Microwave Background Damping Tail}",
  journal = {\apj},
archivePrefix = "arXiv",
   eprint = {1211.4634},
 primaryClass = "astro-ph.CO",
     year = 2013,
    month = feb,
   volume = 764,
      eid = {137},
    pages = {137},
      doi = {10.1088/0004-637X/764/2/137},
   adsurl = {http://adsabs.harvard.edu/abs/2013ApJ...764..137F}
}

@ARTICLE{Zeldovich68,
   author = {{Zeldovich}, Y.~B. and {Kurt}, V.~G. and {Syunyaev}, R.~A.},
    title = "{Recombination of Hydrogen in the Hot Model of the Universe}",
  journal = {Zhurnal Eksperimental noi i Teoreticheskoi Fiziki},
     year = 1968,
    month = jul,
   volume = 55,
    pages = {278-286},
   adsurl = {http://adsabs.harvard.edu/cgi-bin/nph-bib_query?bibcode=1968ZhETF..55..278Z&db_key=AST}
}

@article{DESI2025BAO,
    author = "Abdul Karim, M. and others",
    collaboration = "DESI",
    title = "{DESI DR2 results. II. Measurements of baryon acoustic oscillations and cosmological constraints}",
    eprint = "2503.14738",
    archivePrefix = "arXiv",
    primaryClass = "astro-ph.CO",
    reportNumber = "FERMILAB-PUB-25-0169-PPD",
    doi = "10.1103/tr6y-kpc6",
    journal = "Phys. Rev. D",
    volume = "112",
    number = "8",
    pages = "083515",
    year = "2025"
}

@article{Planck2018VI,
    author = "Aghanim, N. and others",
    collaboration = "Planck",
    title = "{Planck 2018 results. VI. Cosmological parameters}",
    eprint = "1807.06209",
    archivePrefix = "arXiv",
    primaryClass = "astro-ph.CO",
    doi = "10.1051/0004-6361/201833910",
    journal = "Astron. Astrophys.",
    volume = "641",
    pages = "A6",
    year = "2020",
    note = "[Erratum: Astron.Astrophys. 652, C4 (2021)]"
}

@article{Tristram2024,
    author = "Tristram, M. and others",
    title = "{Cosmological parameters derived from the final Planck data release (PR4)}",
    eprint = "2309.10034",
    archivePrefix = "arXiv",
    primaryClass = "astro-ph.CO",
    doi = "10.1051/0004-6361/202348015",
    journal = "Astron. Astrophys.",
    volume = "682",
    pages = "A37",
    year = "2024"
}

@article{Pagano2020,
    author = "Pagano, L. and Delouis, J. -M. and Mottet, S. and Puget, J. -L. and Vibert, L.",
    title = "{Reionization optical depth determination from Planck HFI data with ten percent accuracy}",
    eprint = "1908.09856",
    archivePrefix = "arXiv",
    primaryClass = "astro-ph.CO",
    doi = "10.1051/0004-6361/201936630",
    journal = "Astron. Astrophys.",
    volume = "635",
    pages = "A99",
    year = "2020"
}

@article{ACTDR6PS,
    author = "Louis, Thibaut and others",
    collaboration = "Atacama Cosmology Telescope",
    title = "{The Atacama Cosmology Telescope: DR6 power spectra, likelihoods and {\ensuremath{\Lambda}}CDM parameters}",
    eprint = "2503.14452",
    archivePrefix = "arXiv",
    primaryClass = "astro-ph.CO",
    reportNumber = "FERMILAB-PUB-25-0071-PPD",
    doi = "10.1088/1475-7516/2025/11/062",
    journal = "JCAP",
    volume = "11",
    pages = "062",
    year = "2025"
}

@article{ACTDR6Params,
    author = "Calabrese, Erminia and others",
    collaboration = "Atacama Cosmology Telescope",
    title = "{The Atacama Cosmology Telescope: DR6 constraints on extended cosmological models}",
    eprint = "2503.14454",
    archivePrefix = "arXiv",
    primaryClass = "astro-ph.CO",
    reportNumber = "FERMILAB-PUB-25-0157-PPD",
    doi = "10.1088/1475-7516/2025/11/063",
    journal = "JCAP",
    volume = "11",
    pages = "063",
    year = "2025"
}

@article{LynchA,
    author = "Lynch, Gabriel P. and Knox, Lloyd and Chluba, Jens",
    title = "{Reconstructing the recombination history by combining early and late cosmological probes}",
    eprint = "2404.05715",
    archivePrefix = "arXiv",
    primaryClass = "astro-ph.CO",
    doi = "10.1103/PhysRevD.110.063518",
    journal = "Phys. Rev. D",
    volume = "110",
    number = "6",
    pages = "063518",
    year = "2024"
}

@article{LynchB,
    author = "Lynch, Gabriel P. and Knox, Lloyd and Chluba, Jens",
    title = "{DESI observations and the Hubble tension in light of modified recombination}",
    eprint = "2406.10202",
    archivePrefix = "arXiv",
    primaryClass = "astro-ph.CO",
    doi = "10.1103/PhysRevD.110.083538",
    journal = "Phys. Rev. D",
    volume = "110",
    number = "8",
    pages = "083538",
    year = "2024"
}

@article{KnoxMillea2020,
    author = "Knox, Lloyd and Millea, Marius",
    title = "{Hubble constant hunter{\textquoteright}s guide}",
    eprint = "1908.03663",
    archivePrefix = "arXiv",
    primaryClass = "astro-ph.CO",
    doi = "10.1103/PhysRevD.101.043533",
    journal = "Phys. Rev. D",
    volume = "101",
    number = "4",
    pages = "043533",
    year = "2020"
}

@article{HartChluba2020,
    author = "Hart, Luke and Chluba, Jens",
    title = "{Updated fundamental constant constraints from Planck 2018 data and possible relations to the Hubble tension}",
    eprint = "1912.03986",
    archivePrefix = "arXiv",
    primaryClass = "astro-ph.CO",
    doi = "10.1093/mnras/staa412",
    journal = "Mon. Not. Roy. Astron. Soc.",
    volume = "493",
    number = "3",
    pages = "3255--3263",
    year = "2020"
}

@article{Sekiguchi2021,
    author = "Sekiguchi, Toyokazu and Takahashi, Tomo",
    title = "{Early recombination as a solution to the $H_0$ tension}",
    eprint = "2007.03381",
    archivePrefix = "arXiv",
    primaryClass = "astro-ph.CO",
    reportNumber = "KEK-TH-2238",
    doi = "10.1103/PhysRevD.103.083507",
    journal = "Phys. Rev. D",
    volume = "103",
    number = "8",
    pages = "083507",
    year = "2021"
}

@article{Lee2023,
    author = {Lee, Nanoom and Ali-Ha{\"\i}moud, Yacine and Sch{\"o}neberg, Nils and Poulin, Vivian},
    title = "{What It Takes to Solve the Hubble Tension through Modifications of Cosmological Recombination}",
    eprint = "2212.04494",
    archivePrefix = "arXiv",
    primaryClass = "astro-ph.CO",
    doi = "10.1103/PhysRevLett.130.161003",
    journal = "Phys. Rev. Lett.",
    volume = "130",
    number = "16",
    pages = "161003",
    year = "2023"
}

@article{HyRec,
    author = "Ali-Haimoud, Yacine and Hirata, Christopher M.",
    title = "{HyRec: A fast and highly accurate primordial hydrogen and helium recombination code}",
    eprint = "1011.3758",
    archivePrefix = "arXiv",
    primaryClass = "astro-ph.CO",
    doi = "10.1103/PhysRevD.83.043513",
    journal = "Phys. Rev. D",
    volume = "83",
    pages = "043513",
    year = "2011"
}

@article{Riess2022,
    author = "Riess, Adam G. and others",
    title = "{A Comprehensive Measurement of the Local Value of the Hubble Constant with 1 km s$^{−1}$ Mpc$^{−1}$ Uncertainty from the Hubble Space Telescope and the SH0ES Team}",
    eprint = "2112.04510",
    archivePrefix = "arXiv",
    primaryClass = "astro-ph.CO",
    doi = "10.3847/2041-8213/ac5c5b",
    journal = "Astrophys. J. Lett.",
    volume = "934",
    number = "1",
    pages = "L7",
    year = "2022"
}

@article{CLASS,
    author = "Blas, Diego and Lesgourgues, Julien and Tram, Thomas",
    title = "{The Cosmic Linear Anisotropy Solving System (CLASS) II: Approximation schemes}",
    eprint = "1104.2933",
    archivePrefix = "arXiv",
    primaryClass = "astro-ph.CO",
    reportNumber = "CERN-PH-TH-2011-082, LAPTH-010-11",
    doi = "10.1088/1475-7516/2011/07/034",
    journal = "JCAP",
    volume = "07",
    pages = "034",
    year = "2011"
}

@article{emcee,
    author = "Foreman-Mackey, Daniel and Hogg, David W. and Lang, Dustin and Goodman, Jonathan",
    title = "{emcee: The MCMC Hammer}",
    eprint = "1202.3665",
    archivePrefix = "arXiv",
    primaryClass = "astro-ph.IM",
    doi = "10.1086/670067",
    journal = "Publ. Astron. Soc. Pac.",
    volume = "125",
    pages = "306--312",
    year = "2013"
}

@article{nautilus,
    author = "Lange, Johannes U.",
    title = "{nautilus: boosting Bayesian importance nested sampling with deep learning}",
    eprint = "2306.16923",
    archivePrefix = "arXiv",
    primaryClass = "astro-ph.IM",
    doi = "10.1093/mnras/stad2441",
    journal = "Mon. Not. Roy. Astron. Soc.",
    volume = "525",
    number = "2",
    pages = "3181--3194",
    year = "2023"
}

@article{CosmoPower,
    author = "Spurio Mancini, Alessio and Piras, Davide and Alsing, Justin and Joachimi, Benjamin and Hobson, Michael P.",
    title = "{CosmoPower: emulating cosmological power spectra for accelerated Bayesian inference from next-generation surveys}",
    eprint = "2106.03846",
    archivePrefix = "arXiv",
    primaryClass = "astro-ph.CO",
    doi = "10.1093/mnras/stac064",
    journal = "Mon. Not. Roy. Astron. Soc.",
    volume = "511",
    number = "2",
    pages = "1771--1788",
    year = "2022"
}

@article{Dovekie,
    author = "Popovic, B. and others",
    collaboration = "DES",
    title = "{The Dark Energy Survey supernova program: a reanalysis of cosmology results and evidence for evolving dark energy with an updated Type Ia supernova calibration}",
    eprint = "2511.07517",
    archivePrefix = "arXiv",
    primaryClass = "astro-ph.CO",
    reportNumber = "FERMILAB-PUB-25-0842-CSAID-PPD",
    doi = "10.1093/mnras/stag632",
    journal = "Mon. Not. Roy. Astron. Soc.",
    volume = "548",
    number = "4",
    pages = "stag632",
    year = "2026"
}

@article{Reeves2025,
    author = "Reeves, Alexander and Ferraro, Simone and Nicola, Andrina and Refregier, Alexandre",
    title = "{Multiprobe constraints on early and late time dark energy}",
    eprint = "2510.06114",
    archivePrefix = "arXiv",
    primaryClass = "astro-ph.CO",
    doi = "10.1103/w737-nk6r",
    journal = "Phys. Rev. D",
    volume = "113",
    number = "10",
    pages = "103512",
    year = "2026"
}

@article{Reeves2024,
    author = "Reeves, Alexander and Nicola, Andrina and Refregier, Alexandre and Kacprzak, Tomasz and Valle, Luis Fernando Machado Poletti",
    title = "{12 {\texttimes} 2 pt combined probes: pipeline, neutrino mass, and data compression}",
    eprint = "2309.03258",
    archivePrefix = "arXiv",
    primaryClass = "astro-ph.CO",
    doi = "10.1088/1475-7516/2024/01/042",
    journal = "JCAP",
    volume = "01",
    pages = "042",
    year = "2024"
}

@article{Ong:2025utx,
    author = "Ong, Dily Duan Yi and Yallup, David and Handley, Will",
    title = "{A Bayesian Perspective on Evidence for Evolving Dark Energy}",
    journal = "ArXiv e-prints",
    eprint = "2511.10631",
    archivePrefix = "arXiv",
    primaryClass = "astro-ph.CO",
    year = "2025"
}

@article{Chevallier2001,
    author = "Chevallier, Michel and Polarski, David",
    title = "{Accelerating universes with scaling dark matter}",
    eprint = "gr-qc/0009008",
    archivePrefix = "arXiv",
    doi = "10.1142/S0218271801000822",
    journal = "Int. J. Mod. Phys. D",
    volume = "10",
    pages = "213--224",
    year = "2001"
}

@article{Linder2003,
    author = "Linder, Eric V.",
    title = "{Exploring the expansion history of the universe}",
    eprint = "astro-ph/0208512",
    archivePrefix = "arXiv",
    doi = "10.1103/PhysRevLett.90.091301",
    journal = "Phys. Rev. Lett.",
    volume = "90",
    pages = "091301",
    year = "2003"
}

@article{Peebles1968,
    author = "Peebles, P. J. E.",
    title = "{Recombination of the Primeval Plasma}",
    doi = "10.1086/149628",
    journal = "Astrophys. J.",
    volume = "153",
    pages = "1",
    year = "1968"
}

@article{Seager2000,
    author = "Seager, Sara and Sasselov, Dimitar D. and Scott, Douglas",
    title = "{How exactly did the universe become neutral?}",
    eprint = "astro-ph/9912182",
    archivePrefix = "arXiv",
    reportNumber = "UBC-COS-99-07",
    doi = "10.1086/313388",
    journal = "Astrophys. J. Suppl.",
    volume = "128",
    pages = "407--430",
    year = "2000"
}

@article{ChlubaThomas2011,
    author = "Chluba, J. and Thomas, R. M.",
    title = "{Towards a complete treatment of the cosmological recombination problem}",
    eprint = "1010.3631",
    archivePrefix = "arXiv",
    primaryClass = "astro-ph.CO",
    doi = "10.1111/j.1365-2966.2010.17940.x",
    journal = "Mon. Not. Roy. Astron. Soc.",
    volume = "412",
    pages = "748",
    year = "2011"
}

@article{HartChluba2022,
    author = "Hart, Luke and Chluba, Jens",
    title = "{Varying fundamental constants principal component analysis: additional hints about the Hubble tension}",
    eprint = "2107.12465",
    archivePrefix = "arXiv",
    primaryClass = "astro-ph.CO",
    doi = "10.1093/mnras/stab2777",
    journal = "Mon. Not. Roy. Astron. Soc.",
    volume = "510",
    number = "2",
    pages = "2206--2227",
    year = "2022"
}

@article{Schoneberg2022,
    author = {Sch{\"o}neberg, Nils and Franco Abell{\'a}n, Guillermo and P{\'e}rez S{\'a}nchez, Andrea and Witte, Samuel J. and Poulin, Vivian and Lesgourgues, Julien},
    title = "{The H0 Olympics: A fair ranking of proposed models}",
    eprint = "2107.10291",
    archivePrefix = "arXiv",
    primaryClass = "astro-ph.CO",
    doi = "10.1016/j.physrep.2022.07.001",
    journal = "Phys. Rept.",
    volume = "984",
    pages = "1--55",
    year = "2022"
}

@article{DiValentino2021,
    author = "Di Valentino, Eleonora and Mena, Olga and Pan, Supriya and Visinelli, Luca and Yang, Weiqiang and Melchiorri, Alessandro and Mota, David F. and Riess, Adam G. and Silk, Joseph",
    title = "{In the realm of the Hubble tension{\textemdash}a review of solutions}",
    eprint = "2103.01183",
    archivePrefix = "arXiv",
    primaryClass = "astro-ph.CO",
    reportNumber = "IPPP/20/108",
    doi = "10.1088/1361-6382/ac086d",
    journal = "Class. Quant. Grav.",
    volume = "38",
    number = "15",
    pages = "153001",
    year = "2021"
}

@article{Efstathiou2024,
    author = "Efstathiou, George",
    title = "{Evolving dark energy or supernovae systematics?}",
    eprint = "2408.07175",
    archivePrefix = "arXiv",
    primaryClass = "astro-ph.CO",
    doi = "10.1093/mnras/staf301",
    journal = "Mon. Not. Roy. Astron. Soc.",
    volume = "538",
    number = "2",
    pages = "875--882",
    year = "2025"
}

@article{Poulin2019,
    author = "Poulin, Vivian and Smith, Tristan L. and Karwal, Tanvi and Kamionkowski, Marc",
    title = "{Early Dark Energy Can Resolve The Hubble Tension}",
    eprint = "1811.04083",
    archivePrefix = "arXiv",
    primaryClass = "astro-ph.CO",
    doi = "10.1103/PhysRevLett.122.221301",
    journal = "Phys. Rev. Lett.",
    volume = "122",
    number = "22",
    pages = "221301",
    year = "2019"
}

@article{Colgain2024,
    author = "Colg{\'a}in, Eoin {\'O}. and Sheikh-Jabbari, M. M.",
    title = "{DESI and SNe: Dynamical Dark Energy, $\Omega_m$ Tension or Systematics?}",
    eprint = "2412.12905",
    archivePrefix = "arXiv",
    primaryClass = "astro-ph.CO",
    doi = "10.1093/mnrasl/slaf042",
    journal = "Mon. Not. Roy. Astron. Soc.",
    volume = "542",
    number = "1",
    pages = "L24--L30",
    year = "2025"
}

@article{JedamzikPogosian2020,
    author = "Jedamzik, Karsten and Pogosian, Levon",
    title = "{Relieving the Hubble tension with primordial magnetic fields}",
    eprint = "2004.09487",
    archivePrefix = "arXiv",
    primaryClass = "astro-ph.CO",
    doi = "10.1103/PhysRevLett.125.181302",
    journal = "Phys. Rev. Lett.",
    volume = "125",
    number = "18",
    pages = "181302",
    year = "2020"
}

@article{Chaussidon:2025npr,
    author = "Chaussidon, E. and others",
    title = "{Early time solution as an alternative to the late time evolving dark energy with DESI DR2 BAO}",
    eprint = "2503.24343",
    archivePrefix = "arXiv",
    primaryClass = "astro-ph.CO",
    reportNumber = "FERMILAB-PUB-25-0241-PPD",
    doi = "10.1103/xtql-wh3h",
    journal = "Phys. Rev. D",
    volume = "112",
    number = "6",
    pages = "063548",
    year = "2025"
}

@article{Mirpoorian:2025rfp,
    author = "Mirpoorian, Seyed Hamidreza and Jedamzik, Karsten and Pogosian, Levon",
    title = "{Is dynamical dark energy necessary? DESI BAO and modified recombination}",
    eprint = "2504.15274",
    archivePrefix = "arXiv",
    primaryClass = "astro-ph.CO",
    doi = "10.1088/1475-7516/2025/12/050",
    journal = "JCAP",
    volume = "12",
    pages = "050",
    year = "2025"
}

@article{Reeves:2022aoi,
    author = "Reeves, Alexander and Herold, Laura and Vagnozzi, Sunny and Sherwin, Blake D. and Ferreira, Elisa G. M.",
    title = "{Restoring cosmological concordance with early dark energy and massive neutrinos?}",
    eprint = "2207.01501",
    archivePrefix = "arXiv",
    primaryClass = "astro-ph.CO",
    doi = "10.1093/mnras/stad317",
    journal = "Mon. Not. Roy. Astron. Soc.",
    volume = "520",
    number = "3",
    pages = "3688--3695",
    year = "2023"
}

@article{Gelman:1992zz,
    author = "Gelman, Andrew and Rubin, Donald B.",
    title = "{Inference from Iterative Simulation Using Multiple Sequences}",
    doi = "10.1214/ss/1177011136",
    journal = "Statist. Sci.",
    volume = "7",
    pages = "457--472",
    year = "1992"
}

@article{Poulin:2025nfb,
    author = "Poulin, Vivian and Smith, Tristan L. and Calder{\'o}n, Rodrigo and Simon, Th{\'e}o",
    title = "{Impact of ACT DR6 and DESI DR2 for early dark energy and the Hubble tension}",
    eprint = "2505.08051",
    archivePrefix = "arXiv",
    primaryClass = "astro-ph.CO",
    doi = "10.1103/bx25-1g5d",
    journal = "Phys. Rev. D",
    volume = "113",
    number = "6",
    pages = "063519",
    year = "2026"
}

@article{Lynch:2026pix,
    author = "Lynch, Gabriel P. and Chluba, Jens and Battye, Richard and Vasil, Geoffrey",
    title = "{Limits on the size of electron density fluctuations in the early universe and implications for the Hubble tension}",
    journal = "ArXiv e-prints",
    eprint = "2610.05147",
    archivePrefix = "arXiv",
    primaryClass = "astro-ph.CO",
    year = "2026"
}

@article{Rashkovetskyi:2021rwg,
  title = {Small-Scale Clumping at Recombination and the {{Hubble}} Tension},
  author = {Rashkovetskyi, Michael and Mu{\~n}oz, Julian B. and Eisenstein, Daniel J. and Dvorkin, Cora},
  year = 2021,
  month = nov,
  journal = {Physical Review D},
  volume = {104},
  number = {10},
  eprint = {2108.02747},
  primaryclass = {astro-ph, physics:hep-ph},
  pages = {103517},
  issn = {2470-0010, 2470-0029},
  doi = {10.1103/PhysRevD.104.103517},
  urldate = {2023-08-04},
  archiveprefix = {arXiv},
  langid = {english}
}

@misc{Weiner:2026sfm,
  title = {High-Redshift Physics from the Acoustic Scale},
  author = {Weiner, Zachary J.},
  year = 2026,
  month = mar,
  number = {arXiv:2603.18131},
  eprint = {2603.18131},
  primaryclass = {astro-ph},
  publisher = {arXiv},
  doi = {10.48550/arXiv.2603.18131},
  urldate = {2026-03-21},
  archiveprefix = {arXiv}
}

@article{DESI:2025zpo,
    author = "Abdul Karim, M. and others",
    collaboration = "DESI",
    title = "{DESI DR2 results. I. Baryon acoustic oscillations from the Lyman alpha forest}",
    eprint = "2503.14739",
    archivePrefix = "arXiv",
    primaryClass = "astro-ph.CO",
    reportNumber = "FERMILAB-PUB-25-0167-PPD",
    doi = "10.1103/2wwn-xjm5",
    journal = "Phys. Rev. D",
    volume = "112",
    number = "8",
    pages = "083514",
    year = "2025"
}

\appendix

\section{Adding the local distance ladder}
\label{sec:local_distance_ladder}

\begin{figure*}
\centering
\includegraphics[width=\linewidth]{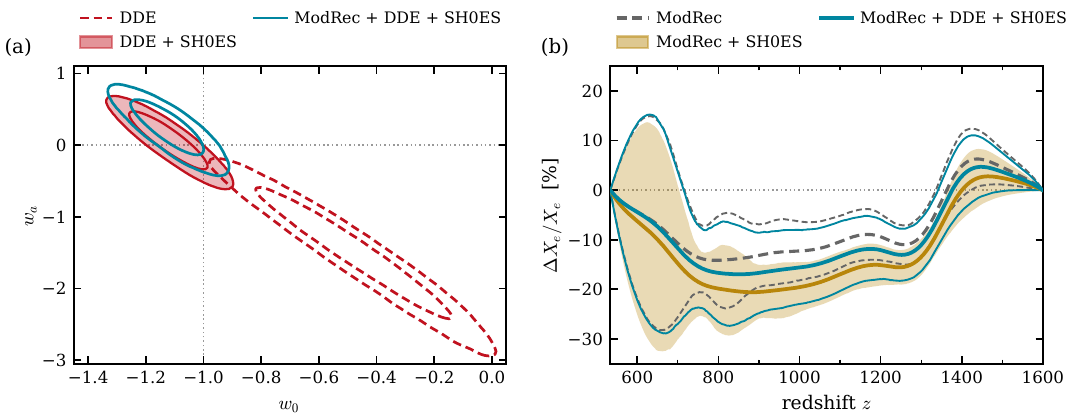}
\caption{Impact of adding the SH0ES distance ladder, for P-ACT. (a) The 68 and 95 per cent credible regions in the $(w_0,w_a)$ plane for DDE before (dashed) and after (filled) adding SH0ES, and for the joint model after adding SH0ES (teal). (b) Reconstructed ionization histories, showing the pointwise posterior mean (thick) and 68 per cent interval (thin, or shaded) for \ModRec\ before (grey, dashed) and after (gold) adding SH0ES, and for the joint model after adding SH0ES (teal). The dotted lines mark \LCDM\ and standard recombination.}
\label{fig:shoes}
\end{figure*}

\begin{table*}
\centering
\caption{\textbf{Impact of adding a SH0ES local distance ladder prior}. The baseline columns give the $H_0$ (in ${\rm km\,s^{-1}\,Mpc^{-1}}$) inferred from the baseline combination of data, without SH0ES, and the tension of that data with SH0ES in each model, in units of $\sigma$. The remaining columns give the increase in the best-fit $\chi^2$ of each model when SH0ES is added, that is, the difference between its best fit with SH0ES and its own best fit without it (importantly this is \textit{not} the difference from \LCDM\ used in Table~\ref{tab:modelcomp}). This total is the sum of two parts: the worsening of the fit to the baseline data (CMB, BAO and \fivetwo), of which the BAO part is explicitly written, and the $\chi^2$ of the SH0ES measurement itself at the new best fit.}
\label{table:shoes}
\setlength{\tabcolsep}{6pt}
\begin{tabular}{llrrrrrr}
\toprule
 & & \multicolumn{2}{c}{Baseline (without SH0ES)} & \multicolumn{4}{c}{Increase in best-fit $\chi^2$ when SH0ES is added} \\
\cmidrule(lr){3-4}\cmidrule(lr){5-8}
CMB data & Model & $H_0$ & Tension with SH0ES & Total & Baseline data & of which BAO & SH0ES \\
\midrule
\textit{Planck} 2018 & \LCDM & $68.4\pm0.3$ & $4.3\sigma$ & $18.4$ & $+1.3$ & $-1.2$ & $17.1$ \\
 & DDE & $63.3\pm1.9$ & $4.1\sigma$ & $16.5$ & $+13.3$ & $+11.1$ & $3.1$ \\
 & \ModRec & $70.3\pm0.7$ & $2.0\sigma$ & $4.1$ & $+1.2$ & $+2.2$ & $2.9$ \\
 & \ModRec$+$DDE & $64.2\pm2.5$ & $2.6\sigma$ & $6.7$ & $+6.0$ & $+6.3$ & $0.7$ \\
\midrule
\textit{Planck} PR4 & \LCDM & $68.4\pm0.3$ & $4.3\sigma$ & $18.3$ & $+1.1$ & $-0.5$ & $17.2$ \\
 & DDE & $63.9\pm2.0$ & $3.7\sigma$ & $13.9$ & $+11.3$ & $+10.4$ & $2.6$ \\
 & \ModRec & $69.1\pm0.7$ & $2.9\sigma$ & $8.7$ & $+2.8$ & $+2.6$ & $5.9$ \\
 & \ModRec$+$DDE & $63.2\pm2.0$ & $3.3\sigma$ & $10.8$ & $+9.7$ & $+7.2$ & $1.1$ \\
\midrule
P-ACT & \LCDM & $68.5\pm0.3$ & $4.2\sigma$ & $18.0$ & $+1.0$ & $-0.9$ & $17.0$ \\
 & DDE & $64.1\pm2.0$ & $3.7\sigma$ & $14.0$ & $+11.2$ & $+10.6$ & $2.8$ \\
 & \ModRec & $69.6\pm0.7$ & $2.6\sigma$ & $7.0$ & $+2.1$ & $+1.8$ & $4.9$ \\
 & \ModRec$+$DDE & $64.2\pm2.3$ & $3.0\sigma$ & $9.1$ & $+8.0$ & $+6.7$ & $1.2$ \\
\bottomrule
\end{tabular}

\end{table*}

In this appendix, we confront each of the models with the value of $H_0$ measured by the SH0ES local distance ladder~\citep{Riess2022}. Before including SH0ES, DDE and \ModRec\ shift the inferred central $H_0$ value in opposite directions with respect to \LCDM\ (Table~\ref{table:shoes}). Considering the combination with P-ACT, \ModRec\ reduces the sound horizon and raises $H_0$ to $69.6\pm0.7\,{\rm km\,s^{-1}\,Mpc^{-1}}$, reducing the discrepancy with SH0ES from $4.2\sigma$ in \LCDM\ to $2.6\sigma$. The DDE model posterior instead gives $H_0=64.1\pm2.0\,{\rm km\,s^{-1}\,Mpc^{-1}}$, lowering the central value but increasing the width of the distribution and hence leaving a $3.7\sigma$ discrepancy similar to the level of tension in \LCDM. The same qualitative conclusions hold for all three CMB likelihoods that we tested (see Table~\ref{table:shoes}).

\begin{figure*}
\renewcommand{\thefigure}{B1} 
\centering
\includegraphics[width=\linewidth]{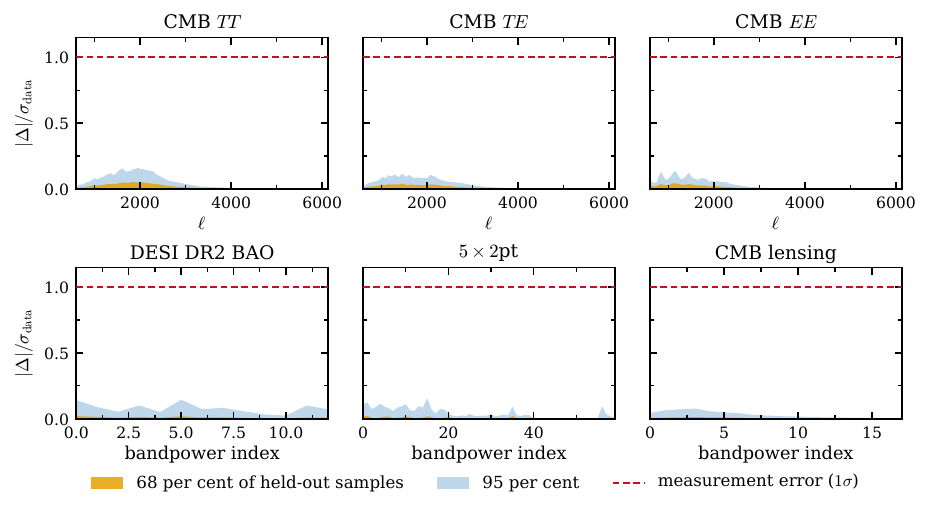}
\\
\caption{\textbf{Emulator errors on held-out samples in units of the corresponding measurement uncertainties.} Top: CMB $TT$, $TE$, and $EE$, normalized by the ACT DR6 bandpower errors. Bottom: DESI DR2 BAO, \fivetwo\ spectra, and CMB lensing. Shaded regions show the 68 and 95 per cent ranges over the held-out samples; the horizontal line marks one standard deviation of the data.}
\label{fig}
\end{figure*}

We then add a Gaussian prior of $H_0=73.04\pm1.04$~\citep{Riess2022} and investigate how this changes the model fits and posteriors. Again focusing on P-ACT, adding SH0ES barely moves the \LCDM\ best-fit, and almost all of the $\chi^2$ increase of $18.0$ is the SH0ES measurement itself ($17.0$). For DDE, adding SH0ES forces the solution away from the expansion history preferred by the baseline data. The posterior shifts towards $w_0\simeq-1.1$ and $w_a\simeq0$ (Fig.~\ref{fig:shoes}a), so that SH0ES itself is fitted well ($\chi^2_{\rm SH0ES}=2.8$), but the $\chi^2$ of the baseline CMB$+$BAO$+$\fivetwo\ data increases by $11.2$. Almost all of this degradation comes from the BAO, whose $\chi^2$ increases by $10.6$, more than twice the improvement of $4.0$ over \LCDM\ that DDE achieves at baseline (Table~\ref{tab:modelcomp}). Thus, DDE cannot accommodate the high local $H_0$ without losing the improvement in the BAO fit that motivates this extension. \ModRec\ accommodates SH0ES more readily by further reducing the sound horizon, which deepens the suppression of the ionization fraction (Fig.~\ref{fig:shoes}b), increasing the $\chi^2$ of the baseline data by only $2.1$. However, the same shift in $a_\star$ that lowers $r_d$ also raises the distance the CMB predicts across the matter era, which the BAO constrain to within $\sim 0.2\%$, so the $H_0$ relief available from recombination is limited by the BAO data. Accordingly, the cost is carried by the BAO, whose $\chi^2$ increases by $1.8$; this exceeds the improvement of $0.9$ at baseline, so the BAO fit becomes worse than that of baseline \LCDM. The joint model behaves similarly: although it can accommodate a higher $H_0$ through recombination freedom, its BAO $\chi^2$ increases by $6.7$. Thus, even with both extensions available, accommodating the local distance ladder comes at the expense of the BAO fit.

These results highlight the different roles of the two extensions. DDE better reproduces the DESI BAO distances but struggles to accommodate the high local $H_0$, whereas modified recombination can raise $H_0$ without reproducing the BAO distance--redshift relation. Neither extension, nor their combination, simultaneously maintains the preferred BAO fit and accommodates SH0ES.

\section{Emulator training and validation}
\label{app:emu}
All theory predictions are obtained with neural-network emulators trained on \textsc{class} outputs over the joint parameter space, following the methodology developed in our previous analyses \citep{Reeves2024,Reeves2025}. We emulate the CMB $TT$, $TE$, and $EE$ spectra to $\ell=6500$, CMB lensing, BAO distances, \fivetwo\ spectra, and supernova distances. The networks are trained on $10^5$--$10^6$ \textsc{class} evaluations, depending on the observable.

Figure~\ref{fig} validates the emulators on independent held-out samples by expressing their prediction error in units of the observational uncertainty of each data vector. Across all probes, the median error remains below $0.03\sigma$, the 68th percentile below $0.06\sigma$, and the 95th percentile below $0.17\sigma$, demonstrating that emulator errors are negligible compared with the measurement uncertainties. The supernova-distance emulator achieves comparable accuracy and is omitted from the figure for clarity.

The CMB emulator training region follows the parameter ranges used in \citet{LynchA,LynchB}, with wider priors in $h$, $w_0$, and $w_a$. All reported posteriors lie well within the high-density region of the training distribution. Only in the CMB-only DDE fits does a small fraction ($4$--$11$ per cent) of the posterior extend beyond the nominal training range in $w_a$; direct comparisons with \textsc{class} in this region give errors consistent with those inside the training range, confirming that this limited extrapolation does not affect our results in these cases. 

\section{Interpolation basis}
\label{app:interp}

\begin{figure*}
\centering
\includegraphics[width=\linewidth]{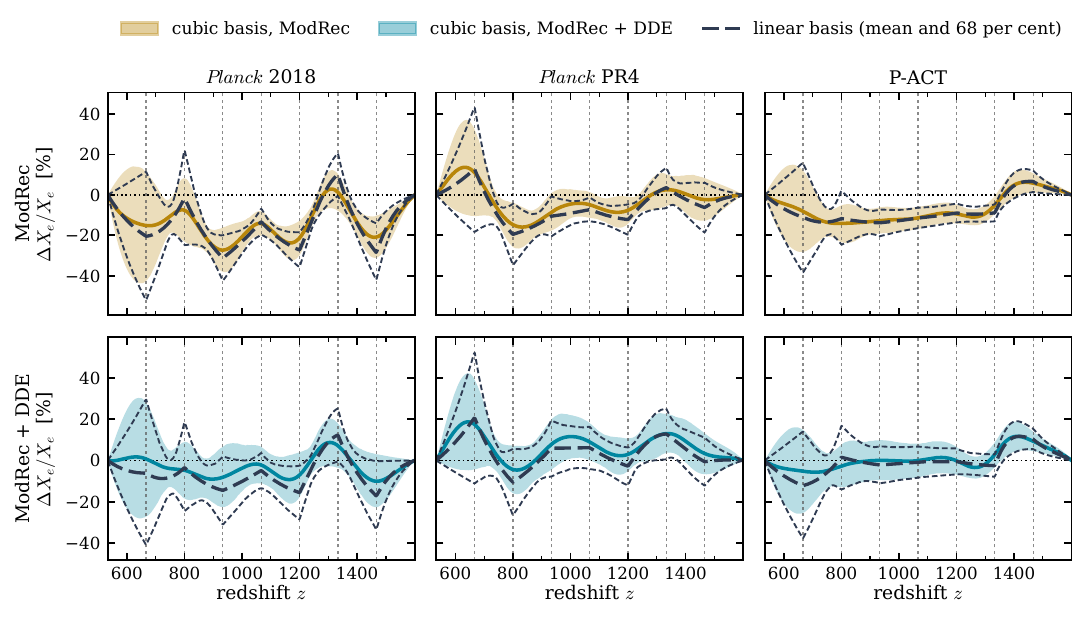}
\\
\caption{\textbf{Impact of the reconstruction basis.} Reconstructed ionization histories using the fiducial clamped cubic basis (filled bands) and a piecewise-linear basis (dark dashed lines), shown for \ModRec{} (gold) and \ModRec$\,+\,$DDE (teal) across the three CMB likelihoods used in this work (columns). Curves and bands show the pointwise posterior means and 68 per cent credible intervals, respectively. The two bases are analysed using independent chains and separately trained emulators.}
\label{fig:linvcub}
\end{figure*}

Several of the reconstructions in Fig.~\ref{fig:joint_reconstruction} oscillate in sign between neighbouring redshifts. While this could be a genuine preference of the data, it is also possible that the cubic-spline interpolation of the node amplitudes could be overshooting between nodes when their amplitudes differ strongly. To test whether this is the case, we repeat the full analysis using piecewise-linear interpolation, which by construction cannot overshoot between the nodes. For this test, we retrained the emulators using the new interpolation scheme. Figure~\ref{fig:linvcub} shows that the reconstructed histories are stable to this choice. The same features appear at the same redshift nodes, and do so with similar amplitudes. The oscillations are therefore preferred by the data rather than being introduced by the interpolation, and are consistent with the poorly constrained higher-order modes identified in principal-component analyses of recombination \citep{Farhang2011,Farhang2013,HartChluba2020}. The preference for modified recombination when $w=-1$, and its disappearance when DDE is allowed, are likewise not affected by this change. The linear basis yields somewhat tighter constraints on the recombination departures in \ModRec, but does not alter their inferred magnitude nor our conclusions.

\bsp
\label{lastpage}
\end{document}